# Optical beam dynamics in a gas repetitively heated by femtosecond filaments

N. Jhajj, Y.-H. Cheng, J. K.Wahlstrand, and H. M. Milchberg*

*Institute for Research in Electronics and Applied Physics, University of Maryland, College Park, Maryland 20742, USA*
*milch@umd.edu

**Abstract:** We investigate beam pointing dynamics in filamentation in gases driven by high repetition rate femtosecond laser pulses. Upon suddenly exposing a gas to a kilohertz train of filamenting pulses, the filament is steered from its original direction to a new stable direction whose equilibrium is determined by a balance among buoyant, viscous, and diffusive processes in the gas. Results are shown for Xe and air, but are broadly applicable to all configurations employing high repetition rate femtosecond laser propagation in gases.

When intense ultrashort pulses propagate through transparent media, they can form filaments, in which the dynamic interplay of nonlinear self-focusing and plasma defocusing lead to an extended core region of high optical intensity and plasma generation [1]. Filamentation has undergone intensive investigation for applications such as harmonic generation [2,3], supercontinuum generation [4], and generation of few-cycle pulses through self-compression

[5,6]. Missing from prior analyses of filamentation are longer timescale effects such as thermal blooming, which has been found to distort the propagation of high power CW lasers in the atmosphere [7, 8]. Despite the high peak power of pulses in femtosecond filaments, the average power in the beam is relatively small, so thermal effects in filamentation are subtle and were only reported recently [9]. In general, models of filamentation had assumed that each laser pulse interacts with a uniform medium and that any perturbations caused by the previous pulse have vanished by the time the next pulse arrives. Recently, it was shown that filamentation in gases at kilohertz repetition rates can be affected by the density depression that accumulates due to laser heating of the gas [9]. Lensing by the density depression alters the propagation of the laser pulse, affecting the onset of filamentation and the spectrum of generated supercontinuum [9].

In this paper, we show that this long-lived density depression, coupled with convective motion of the heated gas, leads to a reproducible deflection of the filamenting beam. We present time-resolved measurements of the accumulating kHz pulse-train-driven gas density depression inside a gas cell and the associated transient beam deflection. We find that the deflection is well-described by a simple model of beam refraction associated with the evolution of the gas density hole, whose time and space scales are well described by a simple fluid analysis. The results presented here provide quantitative understanding of thermal effects on beam propagation of femtosecond filaments, which will point the way to both stabilization and long range control of the filamentation process. Results are shown for Xe and air, but are broadly applicable to all configurations employing high repetition pulse rate femtosecond laser propagation in gases.

Formation of a gas density hole by a single pulse is first briefly reviewed. Depending on the focusing f-number, a variable length plasma is generated either through lens-dominated focusing or by filamentation [1]. Over a timescale depending on the gas type and pressure (~10 ns for air at 1 atm), the plasma then recombines and repartitions most of its initial thermal energy into translational and rotational degrees of freedom of the neutral gas. After ~1 µs the gas reaches pressure equilibrium, and a hot, low density channel occupies roughly the same volume as the original plasma [9]. The gas begins to thermally equilibrate through diffusion, and the density depression expands and fills in over a time scale of milliseconds.

Here, we consider a situation in which a shutter is opened suddenly and a 1 kHz pulse train is focused into a gas cell. The first pulse experiences a uniform gas density. The time separation between successive pulses is short enough that each pulse, which itself heats the gas, is affected by the cumulative density hole, which acts as a negative lens. As the hole evolves, the pulses in the sequence are steered, eventually reaching a steady state deflection. Here we measure and quantify the hole evolution and beam deflection.

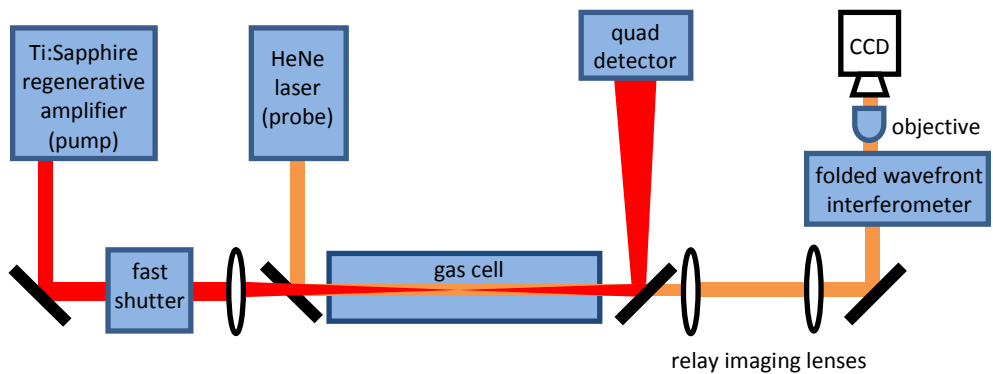

Fig. 1. Experimental setup: the pump and probe travel collinearly through the gas cell. The probe beam is relay imaged at the axial center of the plasma where the density hole is produced.

The experimental setup is shown in Figure 1. A 1 kHz pump pulse train, apertured to 4mm with 45 fs, 120 µJ pulses was focused by a 60 cm lens into a gas cell filled with air or xenon at 1 atm. An additional setup was employed where a 40 cm lens focused the beam into a 2.7 atm gas cell also filled with xenon. The latter is the same setup and gas cell we use for supercontinuum generation for our spectral interferometry experiments [10]. A fast shutter synchronized to the laser system was employed to create a finite pulse train. To capture the transverse location of the pump beam in the far field we routed the output beam, attenuated using neutral density filters, from the gas cell onto a quadrant detector. The beam spot size on the 7.8 mm diameter quadrant detector was ~4 mm. The recorded beam deflections were downward.

Gas density evolution was measured using a folded wavefront interferometer with a CW HeNe laser probe beam [9]. The probe propagated collinearly with the pump pulse train, and was relay imaged from the axial center of the plasma, through a folded wavefront interferometer and onto a CCD. The phase shift from the density perturbation was then extracted from the resulting interferogram using standard interferometric techniques [11]. The phase shift was converted to gas density using the known linear refractive index of Xe [12]. In the case of the 2.7 atm cell, the probe beam propagated at a 5 degree angle with respect to the pump, and Abel inversion was used to extract the size and shape of the density perturbation. Temporal resolution was achieved by time-gating the CCD and triggering the electronic exposure in order to sample the perturbation after a given pulse in the pulse train. The temporal resolution was limited by the minimum CCD exposure duration of ~40 μs, which was short compared to the density profile evolution diffusive timescale as previously measured [9].

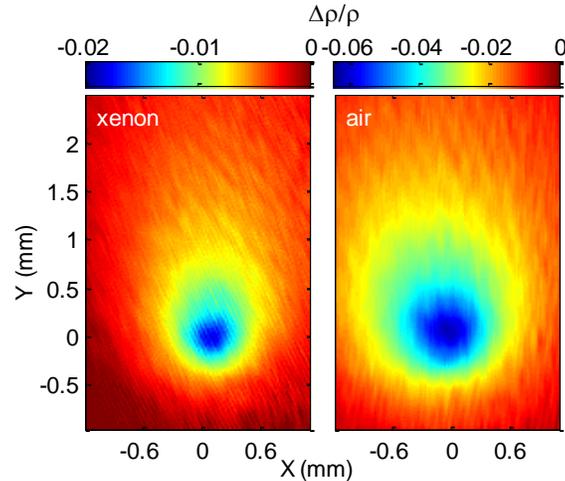

Fig. 2. Two-dimensional plots of gas density in 1 atm xenon and air created by 1 kHz pulse trains with 0.12 mJ and 1.5 mJ per pulse respectively. The image was taken in between pulses in the pulse train. The deep center region was created largely by the pulse prior to the probe, while the larger, shallower background is due to the accumulation of heat from earlier pulses in the train.

A crucial aspect of the gas dynamics not investigated previously is that buoyant forces move the density depression upward. This is seen in Fig. 2, which shows the 2D gas density profile measured in 1 atm xenon and air in response to filament generation at 1 kHz. Here, the filaments are 2.5 cm and 4.75 cm long for xenon and air from visual inspection of the plasma fluorescence. The profile shows the gas density 100 μs before the next pulse in the train. The accumulated density hole from earlier pulses, which has expanded to hundreds of microns in width, is seen to be displaced upward from the pump beam center, as indicated on the image, while the pulse train deflects downward.

Results from the interferometry experiment particularly useful for understanding beam deflection are shown in Fig. 3. Here we use data taken from the 2.7 atm xenon gas cell setup, where the filament is ~3 cm long. Plotted are the hole vertical displacement relative to the pulse train, the hole depth, and the hole half-width-at-half-maximum (HWHM). Each point is an average of 5 measurements. Shot to shot variations were small.

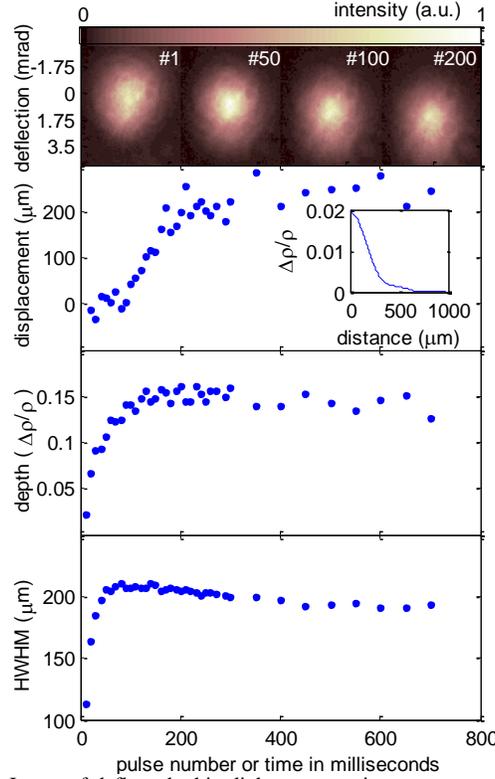

Fig. 3. Top panel: Image of deflected white light supercontinuum spot as a function of pulse number in the 1 kHz pulse train. Lower panels: Hole parameters as a function of pulse number in the pulse train. The fast shutter is opened just before pulse 1. Equilibrium is approached by about the 250[th] pulse (at 250 ms).The inset in the second panel is a typical relative density profile extracted from the interferogram.

The hole displacement was measured by sampling the cumulative density hole just before and after a pulse arrived. The location of the newest pulse is marked by a small, sharp dip in the density superimposed on the larger cumulative density hole, as seen in Fig. 2. The hole depth and HWHM were extracted from an Abel inverted lineout of the interferogram and were measured 100 µs before a pulse arrives. As can be seen from Fig. 2, the hole is significantly more extended above than below its deepest point. Since the hole drifts upwards and the beam deflects downwards, it is the bottom portion of the hole that is relevant to the deflection dynamics, and it is the HWHM of the bottom side of the hole shown plotted in Fig. 3. Also displayed in the top panel of Fig. 3 is a sequence of filament white light beam images in the far field as a function of pulse number in the sequence, showing progressive deflection with negligible beam distortion. Furthermore, we use these beams in our spectral interferometry experiments [10] and detect no phase front distortion from the density hole-induced steering.

We model the beam deflection by the density hole using the ray propagation equation, $\frac{d}{ds}\left(n\frac{d\mathbf{r}}{ds}\right)=\nabla n$, applied in the vertical plane, where $s$ is the optical path along the 2D ray trajectory, $\mathbf{r} = z\hat{\mathbf{z}} + x\hat{\mathbf{x}}$ points to the ray tip, with $\hat{\mathbf{z}}$ along and $\hat{\mathbf{x}}$ vertically perpendicular to the propagation direction, and where $n=n_o+\Delta n$ is the refractive index. For our situation of a small refractive index perturbation $|\Delta n/n_o|<<1$, $n_o\sim 1$ and small ray deflection angles $|dx(z)/dz|<<1$ at large z, this equation simplifies to

$$\frac{d^2x}{dz^2} = \frac{d\Delta n}{dx} \quad (1)$$

The heated density depression relaxes through thermal diffusion [9], and so the index perturbation is approximately Gaussian:

$$\Delta n = \Delta n_0 \exp(-\beta x^2 - \gamma z^2) \quad (2)$$

Here $\beta$ and $\gamma$ characterize the length scales of the perturbation in the $x$ and $z$ directions respectively, and $\Delta n_o$ (<0 for a density hole) gives the maximum amplitude of the perturbation. In the experiment, the axial extent of the density hole ($\propto \gamma^{-1/2}$) of ~ 3 cm was measured by translating the object plane of the imaging system along the axis of the hole, and $\Delta n_o$ and $\beta$ were obtained by fitting the measured density hole to a Gaussian. Light ray trajectories are linear outside of the density hole region, and so the problem can be treated as classical scattering, where the amplitude and dimensions of the index perturbation, impact parameter, and deflection (scattering) angle are the relevant variables. The deflection angle $\Delta\theta$ is given by integrating Eq. (1), using Eq. (2) as the index profile:

$$\left.\frac{dx}{dz}\right|_{z=\infty} - \left.\frac{dx}{dz}\right|_{z=-\infty} = \Delta\theta = \int_{-\infty}^{\infty}\frac{d\Delta n(x(z),z)}{dx}dz \approx \sqrt{4\pi}\Delta n_0 \frac{\beta x_0}{\sqrt{\gamma}}e^{-\beta x_0^2} \quad (3)$$

The incident ray is assumed to be initially travelling parallel to the optical axis with impact parameter $x_o$, where in the experiment $x_o$ corresponds to the density hole displacement.

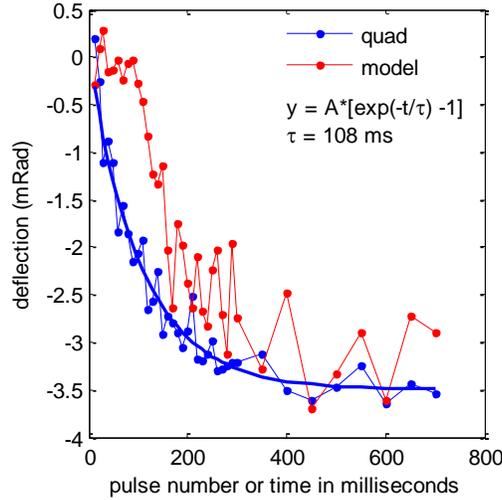

Figure 4. Far field beam deflection angle versus pulse number in 1 kHz plus train. Blue squares: measurement; red squares: ray optics calculation using measured gas density profiles.

Figure 4 shows good agreement between the measured deflection and deflection from the ray optics model. The deviation at earliest times is due to uncertainty in the density hole position for small displacements. The ray optics model contains no free parameters and uses the density hole parameters from Fig. 3 measured simultaneously with the deflection. That this model works so well for filaments is consistent with density hole lensing of the filament 'reservoir,' the lower intensity beam surrounding the filament that exchanges power with the filament core [13]. The beam deflection reaches steady state in ~250 ms with amplitude of ~3.5 mrad. Note that the magnitude of the steady state deflection is reproduced by the model, while the time to steady state agrees with the hole parameter plots of Fig. 3. As can be seen from Fig. 3, the density hole width only increases by a factor of 2 or 3, while the depth and displacement increase by very large factors. The widening of the hole quickly stagnates on the bottom side as convection begins balancing thermal diffusion, preventing heat from spreading downward, as seen in Fig. 2. As a result, the growth in the deflection angle with time is primarily driven by the density hole depth and displacement, and follows their relaxation timescale.

We can estimate the relaxation timescale to steady state using fluid equations in steady state for conservation of momentum and energy. The physical picture is that the gas heated by the laser pulses begins to rise by buoyancy. The gas below the heated region then also rises, setting up a velocity flow field. One can show using the momentum equation that the vertical flow velocity is ultimately limited by viscous drag to $u \approx [\Delta\rho]g\, L_\perp^2/\mu$, where $[\Delta\rho]$ (>0) is the mean decrement in gas density averaged transverse to the flow, $L_\perp$ is the transverse scale length of the density hole, $g$ is the gravitational acceleration, and $\mu$ is the dynamic viscosity. We next consider the energy conservation equation in steady state and apply it to the region just below the localized laser-filament thermal source. Given the vertical flow velocity $u$, the density depression scale length in the vertical direction, $L_\parallel$, adjusts itself until there is a balance between thermal transport and diffusion, giving $[\Delta T]/L_\parallel \sim (^5/_2 N k_B T_o \Gamma u)/\kappa$, where $\rho_o$ and $T_o$ are the background gas density and pressure, $N$ is the gas number density, $[\Delta T]$ (>0) is the transverse average temperature increment, $\Gamma=[\Delta\rho]/\rho_o$ is the relative density depression, $\kappa$ is the gas thermal conductivity, and $k_B$ is Boltzmann's constant. Combining these equations and using pressure equilibrium, $\Delta\rho/\rho_o = \Delta T/T_o$, leads to an effective length scale $L_{eff} = (L_\perp^2 L_\parallel)^{1/3} = \{(\kappa\mu)/(^5/_2 m N^2 \Gamma k_B g)\}^{1/3}$ for the system. Substituting $L_{eff}$ for $L_\perp$ in the expression for $u$ gives $u \approx \{(0.16 mg\kappa^2\Gamma)/(N\mu k_B^2)\}^{1/3}$ for the limiting flow velocity. For a given gas species, note the weak cube root dependence of $u$ and $L_{eff}$ on the experimentally controllable parameters of gas number density $N$ and hole depth $\Gamma$, as well as on the relatively weak temperature dependence of $\kappa$ and $\mu$ [19]. This makes the scale of these estimates robust over a wide range of conditions.

For our experiment in Xe, $\kappa = 5.65 \cdot 10^{-3}$ W·m$^{-1}$·K$^{-1}$ [14], $\mu = 23$ µPa·s [14], $m = 2.2 \cdot 10^{-25}$ kg, $N \sim 7 \cdot 10^{19}$ cm$^{-3}$ at 2.7 atm, and we take a hole depth of $\Gamma \sim 0.10$ from Fig. 3. This gives a limiting gas flow velocity of $u \sim 15$ µm/ms and $L_{eff} \sim 150$ µm. The approximate time to reach this steady state is given by $\Delta t \sim u/a$, where $a = \Gamma g$ is the acceleration from the buoyancy force. Examination of Fig. 3 shows that in the early phase of hole displacement, $\Gamma \sim 0.01$-$0.05$, giving $\Delta t \sim 20$-$100$ ms, a range of timescales in reasonable agreement with the approach to steady state shown in Fig. 3 and Fig. 4.

In an alternative, essentially kinematic approach for estimating these space and time scales, we take an estimate of the initial height rise of the hole due to buoyancy, $d_{rise} \sim \frac{1}{2} \Gamma g t^2$ and set it equal to the approximate hole radius as determined by thermal diffusion, $d_{thermal} \sim 2(\alpha t)^{1/2}$ [9], the idea being that the thermal source location cannot move below the expanding hole width as the hole rises. Here, $t$ is the elapsed time after the laser shutter is opened and $\alpha = \kappa/(^5/_2 N \cdot k_B)$ is the thermal diffusivity. The result is $t = t_{rise} \sim (4/\Gamma g)^{2/3} \cdot \alpha^{1/3}$ and $d_{rise} \sim (32 \alpha^2/\Gamma g)^{1/3}$. Using the above parameters gives $t_{rise} \sim 100$ ms and $d \sim 1$ mm, in reasonable agreement with the experiment and the prior scale estimates.

In summary, we have examined the propagation of a high repetition rate filament pulse train in the transiently evolving gas that it heats. When a filament pulse train is suddenly initiated in a gas, the local density is reduced, driving buoyant motion of the gas. The upward drift of the gas density hole steers subsequent pulses downward. Eventually the buoyant motion is damped by viscous forces, establishing a steady gas flow field and hole density profile, and the downward beam steering stabilizes. Unlike conventional thermal blooming with high power CW lasers, the beam mode is preserved. Further, a simple ray model explains the transient deflection of the beam by the accumulated density hole, and simple fluid analysis explains the experimentally observed space and time scales for the density hole dynamics and beam steering.

**Acknowledgements**
The authors thank Sina Zahedpour and David Meichle for useful discussions. This work is supported by the Air Force Office of Scientific Research, the Office of Naval Research, the National Science Foundation, and the US Dept. of Energy.